\documentclass[10pt, twoside, twocolumn]{article}
\usepackage{bm}
\usepackage{Latex-document}
\usepackage[usenames]{color}
\definecolor{lightblue}{rgb}{0, 0, 0}

\usepackage{latexsym}
\usepackage{amsfonts}
\usepackage{amsmath}
\usepackage{amssymb}
\usepackage{color}
\usepackage{epsfig}
\usepackage{xspace}
\usepackage{subfigure}
\usepackage{graphicx}
\usepackage{balance}
\usepackage{cite}

\usepackage[english]{babel}

\usepackage{epstopdf}
\usepackage{enumerate}
\usepackage{color}

\newcommand{\eat}[1]{}

\newcommand{\bi}{\begin{itemize}}
\newcommand{\ei}{\end{itemize}}

\newcommand{\be}{\begin{enumerate}}
\newcommand{\ee}{\end{enumerate}}
\newcommand{\beqn}{\begin{eqnarray*}}
\newcommand{\eeqn}{\end{eqnarray*}}

\newcommand{\stitle}[1]{\vspace{1ex}\noindent{\bf #1}}

\newcommand{\ie}{\emph{i.e.,}\xspace}
\newcommand{\eg}{\emph{e.g.,}\xspace}
\newcommand{\wrt}{\emph{w.r.t.}\xspace}

\newcommand{\kw}[1]{{\ensuremath {\mathsf{#1}}}\xspace}

\newcounter{ccc}

\newcommand{\CFDconsistency}{{\mbox{\s
\documentclass{sig}
\usepackage{algorithm}
\usepackage[noend]{algorithmic}
\usepackage{latexsym}
\usepackage{amsfonts}
\usepackage{amsmath}
\usepackage{amssymb}
\usepackage{color}
\usepackage{epsfig}
\usepackage{xspace}
\usepackage{graphicx}
\usepackage{times}
\usepackage{subfigure}mall\sf CFD\_Checking}\xspace}}

\newcommand{\NP}{{\sc np}\xspace}

\newcommand{\E}{{\cal E}}
\newcommand{\eop}{\hspace*{\fill}\mbox{$\Box$}}     
\newcounter{example}
\renewcommand{\theexample}{\arabic{example}}

\newcommand{\nthesection}{\arabic{section}}
\newcounter{theorem}
\renewcommand{\thetheorem}{\arabic{theorem}}
\newcounter{prop}
\renewcommand{\theprop}{\arabic{theorem}}
\newcounter{lemma}
\renewcommand{\thelemma}{\arabic{theorem}}
\newcounter{cor}
\renewcommand{\thecor}{\arabic{theorem}}
\newcounter{definition}[section]
\renewcommand{\thedefinition}{\nthesection.\arabic{definition}}

\newcounter{alg}[section]
\renewcommand{\thealg}{\nthesection.\arabic{alg}}

\newcounter{arule}
\renewcommand{\thearule}{\arabic{arule}}

\newcounter{claim}
\renewcommand{\theclaim}{\arabic{claim}}

\renewcommand{\texttt}[1]{{\small\textsf{#1}}}

\definecolor{gray}{rgb}{0.5,0.5,0.5}

\newcommand{\titlename}{$\bm{Big~Graph~Search:~Challenges~and~Techniques}$}
\newcommand{\authorname}{Shuai Ma, Jia Li, Chunming Hu, Xuelian Lin, Jinpeng Huai}

\newcommand{\atime}{Received month dd.yyyy; accepted month dd.yyyy}
\newcommand{\aemail}{\{mashuai, lijia, hucm, linxl, huaijp\}@buaa.edu.cn}

\date{}

\begin{document}

\thispagestyle{first}
\setcounter{page}{1}

\begin{tabular*}{\textwidth}{l}
 \hspace*{-6.1mm}\includegraphics{xian.eps}\vspace{-6.4mm}\\
 \hspace*{-6mm}\colorbox{lightblue}{
\arraycolsep=132pt \normalsize\hspace*{-10mm}{\color[cmyk]{.0, 0.0, 0, .0}$\begin{array}{l}\\[-3.5mm]\bf \hspace*{-37mm}
REVIEW~ARTICLE
\end{array}$}}\vspace{-2mm}\\
\end{tabular*}

\begin{strip}
\begin{center}
{\tfont \LARGE \titlename}\\[6mm]
{\bf \authorname}\\[3mm]
\normalsize{State Key Laboratory of Software Development Environment\\School of Computer Science and Engineering  \\Beihang University, Beijing 100191, China}\\
\end{center}
\cnote
\end{strip}

\Abstract{
On one hand, compared with traditional relational and XML models, graphs have more expressive power and are widely used today.  On the other hand, various applications of social computing trigger the pressing need of a new search paradigm. In this article, we argue that big graph search is the one filling this gap. To show this, we first introduce the application of graph search in various scenarios. We then formalize the graph search problem, and give an analysis of graph search from an evolutionary point of view, followed by the evidences from both the industry and academia. After that, we analyze the difficulties and challenges of big graph search. Finally, we present three classes of techniques towards big graph search: query techniques, data techniques and distributed computing techniques.

\eat{
give a try of formalizing the concept of graph search. Then we give an analysis of graph search from an evolutionary point of view and point out its urgent need, followed by the evidences from both the industry and academia.

n this article we have introduced the challenges and opportunities of big graph search, a new promising paradigm for social computing in the big data era. Firstly we have analyzed the need of big graph search with various applications, industrial and academia developments, and the evolution history of information searching paradigms. Second, we have pointed out the challenges of big graph search. Finally, we have introduced three types of techniques towards big graph search: query, data and distributed computing techniques.

soc has brought about the emergence of graph search, a new paradigm
for social computing, which has drawn more and more attention from both industry and academia.
Graph search has found its applications in biology, chemistry, social network, traffic route selection,
recommender systems and so on. In this article, firstly, we will focus on the applications of graph search in
industry and the research development in academia, and then reveal the significance and necessity of graph search
in many occasions subsequently.
}

}

\footnote{\footname}


---------------------
\vspace{-0ex}
\section{Introduction}
\label{sec-intro}

With the rapid development of social computing, Internet and various applications have brought about exponentially growing data. According to the recent report of the UN's International Telecommunications Union (ITU), Internet users will hit 3 billion globally by the end of 2014~\cite{itu};  The total number of monthly active Facebook users has reached over 1.3 billion, and the increment of its users from 2012 to 2013 is about 22\%~\cite{facebooknum}. All these indicate the coming of  an era of big data. Indeed, ``data are becoming the new raw material of business: an economic input almost on a par with capital and labour.''~\cite{dataevery}.
How to filter unnecessary data and find the desired information so that one could easily make timely and accurate decisions? This has become one of the most pressing needs in such a big data era.

Compared with traditional relational and XML models, graphs have more expressive power, and play an important role in many applications, such as social networks, biological data analyses, recommender systems, complex object identification and software plagiarism detection.   Essentially, this is because the core data involved in these applications can be conveniently represented as graphs. For instance,  a social network (\eg Facebook~\cite{facebook}, Twitter~\cite{twitter} and Weibo~\cite{weibo}) constitutes all kinds of social users/activities, which is essentially a graph, whose nodes denote  users/activities and edges denote their relationships, such as friendships, respectively.

The wide use of graphs has brought about the emergence of big graph search, \ie retrieving information from big graphs in a timely and accurate manner, which has drawn more and more attention from both the industry and academia~~\cite{cccf-1,cccf-2,ict-3}. We first give an overview of the application scenarios of graph search.

\eat{Internet has become rather an important way for people to acquire knowledge~\cite{socialhttp}. And with the development of emerging technologies such as wireless Internet and \textcolor{red}{3G(4G?)}, people communicate with each other more than before by using BBS, Facebook and Twitter etc. We could also find that}

\stitle{(1) Social networks and the Web.} Nowadays,  the rapid development of the Web and social networks has made significant influences on people's social and personal behaviors. Take for instance, Facebook: (a) the total number of its users is very large: there are more than 1.3 billion monthly active users and 0.68 billion mobile users till June 2014; (b) the relations among users and other objects are tight: a user has 130 friends and likes 80 pages on average; (c) there is a large amount of information dissemination on Facebook: more than 4.75 billion pieces of content are shared daily; (d) the site visit of Facebook is quite frequent: 23\% of users check Facebook 5 times or more daily, and  a user spends 20 minutes on the site per visit on average~\cite{facebooknum}.

As mentioned earlier, social networks can be easily represented by graphs, which comes with all kinds of graph search techniques~\cite{TianP08,FanLMTWW10,Barcelo10,FengCBM14}, including neighbor query and social network compression~\cite{MaserratP10}. Similar to social networks, the Web can be expressed as a big graph as well, whose nodes denote Web pages, and edges indicate hyperlink relationships between Web pages. In fact, the Web site classification and Web mirror detection problems  can be treated as the graph classification~\cite{SchenkerLBK04} and  graph matching problems~\cite{FanLMWW10}, respectively.

\begin{figure}[tb!]
  \begin{center}
  \includegraphics[scale=0.56]{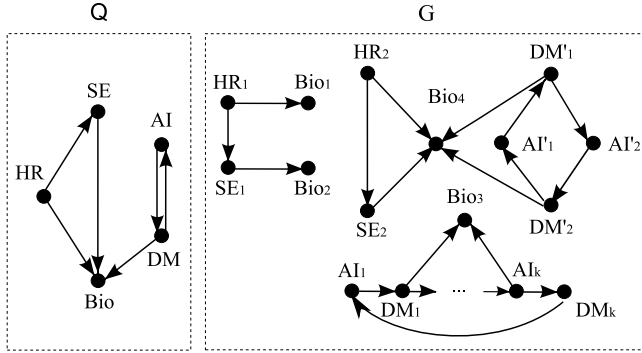}\\
  \end{center}
  \vspace{-3ex}
  \caption{A recommendation network}\label{fig-recommendation}
  \vspace{-3ex}
\end{figure}

\stitle{(2) Recommender systems.} Recommendation has found its usage in many applications, such as social matching systems, and graph search is a useful tool for recommendation~\cite{TerveenM05}. Consider the example that a headhunter wants to find a biologist (Bio) to help a group of software engineers (SEs) analyze genetic data~\cite{ShuaiMaVLDB12,tods-MaCFHW14}. To do this, she uses an expertise recommendation network $G$, as depicted in Fig.~\ref{fig-recommendation}, in which nodes denote persons labeled with their expertise, and edges indicate recommendations, e.g. $\text{HR}_1$ recommends $\text{Bio}_1$, and $\text{AI}_1$ recommends $\text{DM}_1$. The biologist Bio needed is specified with a pattern graph $Q$, also shown in Fig.1. We could find that Bio has to be recommended by: (a) an HR, an SE and a data mining expert (DM) together, as data mining knowledge is required for the job, (b) the SE is also recommended by the HR, and (c) there is an artificial intelligence expert (AI) who recommends the DM and is recommended by the DM. Based on the pattern graph $Q$ and data graph $G$, the headhunter could find the suitable biologist in $G$ who meets the requirements, by utilizing graph search techniques developed in~\cite{ShuaiMaVLDB12,tods-MaCFHW14}.

\stitle{(3) Complex object identification.} Data quality problem costs U.S. business more than \$600 billion a year~\cite{WWEckerson02}, and data cleaning techniques can help mitigate the losses to a large extent, \eg it delivers an overall business value of more than ``600 million GBP'' each year at BT by adopting data cleaning tools~\cite{BOtto09}. Data cleaning  typically contains two central issues: record matching and data repairing~\cite{FanRMDR11}. Complex object identification is the most difficult issue in record matching, which is to identify complex objects referring to the same entity in a physical world. One possible solution is to represent complex objects as graphs, and then to identify the same ones by utilizing graph search techniques, such as subgraph isomorphism and graph homomorphism~\cite{FanLMWW10,Ullmann76}.

\stitle{(4) Software plagiarism detection.} With the popularity of open-source software, it gets much easier for a less self-disciplined developer to use (part of) other software without giving proper credits. Traditional plagiarism detection tools are not adequate for finding serious software plagiarism cases.  A novel plagiarism detection tool has been developed based on graph search techniques~\cite{LiuCHY06}. Firstly, it transforms the source and target programs into program dependence graphs~\cite{FerranteOW87}. Secondly, it tests the similarity of the two  program dependence graphs with subgraph isomorphism~\cite{Ullmann76}. Finally, if the graph similarity is high enough, it concludes the plagiarism. The rational behind this is that the core structure and control flow of programs, reflected by their program dependence graphs, are hardly to be modified.

\begin{figure}[tb!]
  \begin{center}
  \includegraphics[scale=0.35]{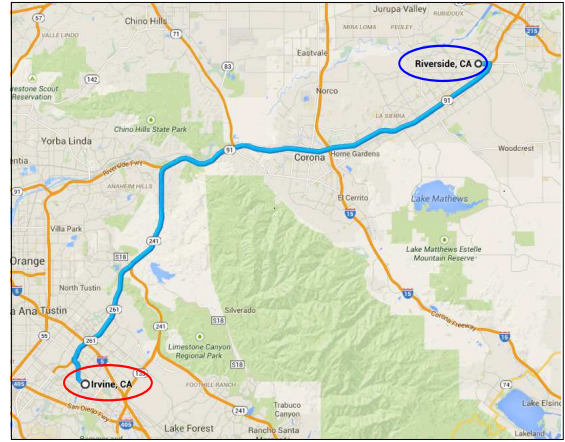}\\
  \end{center}
  \vspace{-3ex}
  \caption{A route planning example}\label{fig-routeexample}
  \vspace{-3ex}
\end{figure}

\stitle{(5) Traffic route planning.} Graph search is a common practice in transportation networks, due to the wide application of location-based services. Consider an example taken from \cite{RiceT10}. Mark is a driver in  the U.S. who wants to travel from Irvine to Riverside in California. (a) If Mark wants to reach Riverside by his car in the shortest time, this can be treated as the classical shortest path problem~\cite{CormenLRS01}, based on which Mark can figure out his best solution from Irvine to Riverside is by traveling along State Route 261, as illustrated by Fig.~\ref{fig-routeexample}. (b) However, if Mark drives a truck carrying with hazardous materials, which may not be allowed to cross over some bridges or railroad crossings, then a pattern graph approach specifying route constraints with regular expressions may be needed to find an optimal transport route~\cite{Zaiben09}.

In addition, graph search techniques have also been adopted  in virtual networks~\cite{ChowdhuryRB09}, pattern recognition~\cite{ConteFSV04} and VLSI design~\cite{KarypisAKS99}, among other things.


\begin{figure*}[tb!]
  \vspace{1.5ex}
  \begin{center}
  \includegraphics[scale=0.68]{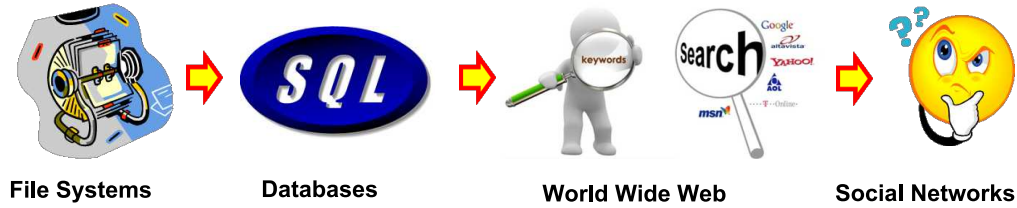}\\
  \end{center}
  \vspace{-4.5ex}
  \caption{The evolution of  information search}\label{fig-importance}
  \vspace{-3ex}
\end{figure*}

\stitle{Organization}. In the rest of this article, we first give a formal definition of graph search and explain why it is important in Section~\ref{sec-definition}. Then we introduce the challenges of big graph search in Section~\ref{sec-problems}, followed by techniques towards big graph search in Section~\ref{sec-tech}. Finally, we conclude  in Section~\ref{sec-conclusion}.

\vspace{-0ex}

\section{Graph Search, Why Bother?}
\label{sec-definition}

In this section, we first give a try of formalizing the concept of graph search. Then we give an analysis of graph search from an evolutionary point of view and point out its urgent need, followed by the evidences from both the industry and academia.

\subsection{What is Graph Search}
\label{subsec-def}

 We first formalize the concept of graph search.

\stitle{Graph search}. Given two graphs $G_p$, also referred to as the pattern graph and $G_d$, also referred to as the data graph, graph search is (1)  to  decide whether $G_p$ ``matches'' $G_d$, or (2) to identify the subgraphs of $G_d$ that $G_p$ ``matches''.

Here graphs consist of nodes and edges, both of which are often attached with labels indicating all kinds of information.
Pattern graphs are usually small, \eg with several or dozens of nodes/edges, while data graphs are often big, \eg with billions of nodes/edges.

Graph search covers two classes of queries: (1) the first class is boolean queries, \ie to answer ``yes'' or ``no'', and (2) the second one is functional queries, \ie to identify and return the matching subgraphs. It is obvious that functional queries may need the aid of boolean queries.

\stitle{Remarks}. The above definition of graph search is quite general, as different semantics of ``match'' lead to different graph search queries~\cite{cccf-1,cccf-2}.
Most, if not all, common graph queries belong to graph search queries, such as node queries (\eg neighbor query~\cite{MaserratP10}), path queries (\eg reachability~\cite{FanLMTW11} and shortest path~\cite{CormenLRS01}) and subgraph queries (\eg graph homomorphism~\cite{FanLMWW10}, subgraph isomorphism~\cite{Ullmann76}, graph simulation~\cite{FanLMTWW10} and its extension strong simulation~\cite{ShuaiMaVLDB12}).

\subsection{An Evolutionary Point of View}

A serious question arises naturally: why do we need another search paradigm -- graph search? We next answer this question from an evolutionary point of view.

\eat{why graph search and related techniques in recent years have drawn more and more attention from both industry and academia?
Social network emerges in social computing era, and in history, every time an important object emerges would lead to the change in searching paradigm. It is necessary for us to make a review on the historical development of searching paradigms, depending on which we could answer two questions: (1) why we need a new searching paradigm to the social computing era and (2) what kind of searching paradigm do we need now?

which has undergone a serious change from  file system, to .

As shown in Fig.~\ref{fig-importance}, searching approaches adopted by computers have undergone such a process: File system search $\rightarrow$ Database system search $\rightarrow$ Web Network search $\rightarrow$ Social Network search.
}

Consider the evolution roadmap of information search shown in Fig.~\ref{fig-importance}. The emphasis of information search has undergone a serious shift, \ie from file systems, to database systems, to the World Wide Web, and to the most recent social networks:

\bi
\item {\em File systems}. Since the 1960s, computers have been equipped with modern operating systems~\cite{Hansen01}.  The file system in an operating system is an abstraction to store and organize a set of computer files, and it usually supports users to look for specific files, \ie simple searching functionalities.

\item {\em Database systems}. In the mid-1960s, database systems began being applied in business, and, subsequently, relational databases played a dominant role.  Since the late 1970s, the invention of structured query language (SQL) has significantly promoted the use of databases~\cite{Raghu2000}.

\item {\em The Web}. In the 1990s,  search engines, such as Google, Bing and Yahoo!, are widely used due to the blossom of the World Wide Web.
These search engines unanimously adopted the simple but very useful approach --  keyword search, which provides people with a convenient and easy way to search specific information on the Web.

\item {\em Social networks}. From the end of last century, with the rising of Web 2.0, social networks have made significant influences on the society.
However, a dominant search paradigm seems missing in such an era of social computing and big data.
\ei
\vspace{-1ex}

As the above analysis shows,  an important IT invention, \eg file systems, database systems and the Web, usually triggers the emergence of a novel search paradigm. We are essentially in a situation to look for one for social computing and social networks, and we believe that {\em graph search is the one filling the  gap}. The ``graph search''~\cite{FbGraphsearch} and ``knowledge graph''~\cite{GoogleKGraph1} released by Facebook  and Google, respectively, shed light on this. However, another question arises: why could not we simply use SQL or keyword based search?

\stitle{(1) Graph search vs. SQL search}. SQL search is a very strong supporting tool for searching information from relational database systems. However, it is not appropriate for  searching information from graphs even though graphs could be stored using relations, due to its disability and inconvenience for answering recursive queries such as graph reachability and shortest paths~\cite{AbHuVi1995}. Indeed, for simple graph queries that SQL search would do, graph search could do even better. We next illustrate this with an example taken from~\cite{Sakr2011}, a simple searching case of ``finding the names of all of Alberto Pepe's friends in a social network''.

\begin{figure}[tb!]
  \begin{center}
  \includegraphics[scale=0.4]{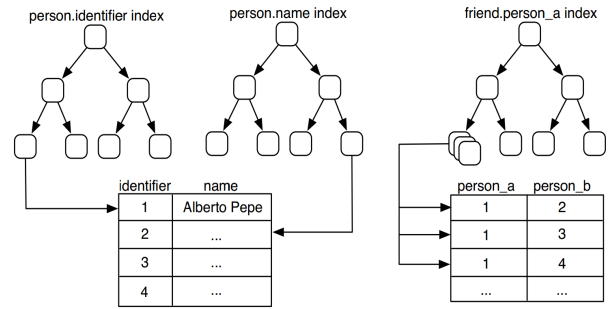}\\
  \end{center}
  \vspace{-3ex}
  \caption{Relational representation}\label{fig-relationgraphcmp1}
  \vspace{-3ex}
\end{figure}

\noindent{\em Case 1: Social networks are stored using relations.}
There are two relations: \kw{person(identifier, name)} for storing a person's unified identifier and its name, and \kw{friend(person\_a, person\_b)} for storing the friendship of persons with identifiers \kw{person\_a} and \kw{person\_b}.
In addition, two {$B^+$--$tree$} indexes are built on each column of the \kw{person} relation: the person.identifier and person.name indexes, and one index is built on the \kw{person\_a} column of the  \kw{friend} relation: the friend.person\_a index. We assume that there are in total {\em n} persons and {\em m} friendships.
The relational representation is presented in Fig.\ref{fig-relationgraphcmp1}.

To get the names of Alberto Pepe's friends, three steps are necessary, as shown in the following.

\bi
  \item[(a)] Find the unique identifer of ``Alberto Pepe'' from relation \kw{person}, which takes $O(\log_{2} n)$ time  using the person.name index.

  \item[(b)] Find all the $k$ identifiers of the friends of ``Alberto Pepe'' from relation \kw{friend} with the identifer found in (a), which takes   $O(\log_{2}n + k)$ time using  the friend.person\_a index.


   \item[(c)] Find the $k$ friends' names from relation \kw{person} with the $k$ identifiers found in (b), which takes $O(k\log_{2}n)$ time using the person.identifier index.
\ei
\vspace{-1ex}

\begin{figure}[tb!]
  \begin{center}
  \includegraphics[scale=0.4]{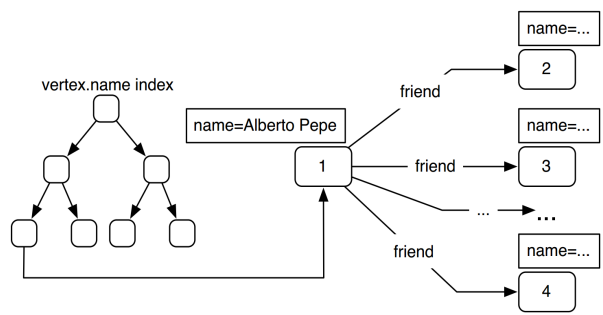}\\
  \end{center}
  \vspace{-3ex}
  \caption{Graph representation}\label{fig-relationgraphcmp2}
  \vspace{-3ex}
\end{figure}

\noindent{\em Case 2: Social networks are stored using naive graphs}.
The person and friendship information can be stored as a graph as shown in Fig.~\ref{fig-relationgraphcmp2}. Each person can be represented as a node labeled with the person's name and unified identifier, and the friendship between two persons can be represented as an edge between the two corresponding nodes. A  {$B^+$--$tree$}  index is built on the graph, vertex.name index, to quickly locate the position of a node in the graph with a person's name.

To get the names of Alberto Pepe's friends, two steps are needed, as shown below.

\bi
  \item[(a)] Identify the node with name ``Alberto Pepe'', which takes $O(\log_{2}n)$ time using the vertex.name index.
  \item[(b)] Find the $k$ friend nodes of the node found in (a) by traversing its adjacent neighboring nodes and get the friend names directly in the $k$ node labels, which takes $O(k+y)$ time such that $k+y$ is the total number of the neighboring nodes.
\ei
\vspace{-1ex}

 It is obvious that the searching speed  is improved from $O((k+2)\log_{2}n)$ to $O(\log_{2}n)$ when using the graph representation, instead of the relational representation. The improvement is crucial when $n$ is really large, \eg when there are billions of users. Of course, one could add redundant information to speed up its efficiency, which results in extra space cost in turn. Hence, for big graph search, the graph search approach is much superior to the SQL search approach.

\stitle{(2) Graph search vs. keyword search}. The traditional keyword based searching approach is mainly for retrieving information from the Web, which is not appropriate for searching information from social networks. The information on the Web is usually isolated and {\em object--object weak tied} from each other, and mainly about ``historical and existing'' information, \ie what happened and happening. Social computing generally takes the social factors into consideration, such as the social structure, organization and activity, which makes {\em relations} a dominant role in social search. Besides, social data are usually {\em person--person strong related}  or {\em person--object strong related}. This makes the {\em future and relation} information particularly important for social search. Under these circumstances, the keyword based searching approaches cannot meet the requirements raised by social computing and social networks nowadays.

Hence we argue that  graph search is a new searching paradigm for social computing in the big data era. Indeed, Facebook has provided a new searching technique named ``Graph Search''~\cite{FbGraphsearch}, which allows users to search for information using simple natural language sentences, \eg ``Restaurants in New York that my friends like'', ``Photos taken in Hawaii of my friends'' and ``National parks where my friends have been to''.
Besides, the development of social networks has also promoted the urgent need of a new search engine in turn.

\subsection{Joint Efforts of the Industry and Academia}

\eat{Also mention the research teams from Social Networks and Graph Matching, Communications of CCF, Volume 8, Number 4, 20-24, 2012. \cite{cccf-2} \cite{cccf-1} also mention Facebook and Google.}

\begin{figure}
  \begin{center}
  \includegraphics[scale=0.52]{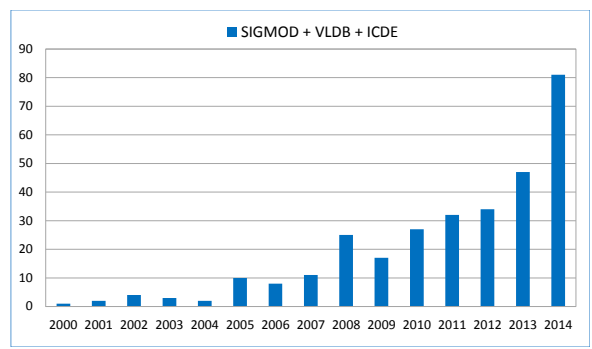}\\
  \end{center}
  \vspace{-3ex}
  \caption{Statistics of papers on graphs}\label{fig-paperstatistics}
  \vspace{-3ex}
\end{figure}

Recently, we have conducted a survey on the number of papers on graphs published in the top three influential database system conferences (SIGMOD, VLDB and ICDE) ever since 2000. The result is shown in Fig.~\ref{fig-paperstatistics}, from which we have found that: (a) from around 2000 (the emergence of Web 2.0), researchers began to focus on the study of graphs, (b) the number of papers on graphs has been increasing continuously since then, (c) from 2008, graphs have been a hot topic in the field of database research, and (d) there is a significant increment of the number of papers on graphs in 2014.

Many well-known research institutions and companies have been concentrating on the research and applications of graphs. For example, Microsoft's Trinity~\cite{microsofttrinity} project and ``Horton - Querying Large Distributed Graphs''~\cite{microsoftldg} project for data center; large-scale graph processing system Pregel~\cite{MalewiczABDHLC10} of Google; ``Knowledge Acquisition and Management''~\cite{yahoopartition} project of Yahoo!; Neo4j's open-source graph database~\cite{Neo4j};  ``Graph Search'' of Facebook~\cite{FbGraphsearch}; and the research teams from academia such as the University of California Santa Barbara,  University of Edinburgh, University of New South Wales, Chinese University of Hong Kong, and Beihang University.

The joint interests and efforts from both the industry and academia provide more evidences on the power and importance of graph search.

\vspace{-0ex}
\section{Challenges of Big Graph Search}
\label{sec-problems}

In this section, we first introduce the \kw{FAE} rule that is important for a search engine, and we then point out its difficulties and challenges for big graph search.

\subsection{The \kw{FAE} Rule}

\begin{figure}
  \begin{center}
  \includegraphics[scale=0.4]{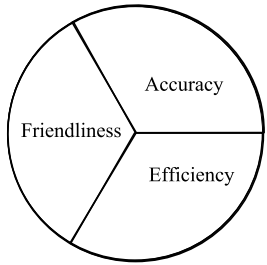}\\
  \end{center}
  \vspace{-3ex}
  \caption{The FAE rule}\label{fig-faerule}
  \vspace{-3ex}
\end{figure}

The \kw{FAE} rule says that the quality of search engines involves with three key factors: {\em friendliness}, {\em accuracy} and {\em efficiency}, as illustrated in Fig.~\ref{fig-faerule}, and that a good search engine must provide the users with a friendly query interface and highly accurate answers in a fast way.

\stitle{(1) Friendliness}. It is necessary for a search engine to provide the users with a friendly query interface such that the users could conveniently specify their searching conditions with small efforts.

Generally speaking, the keyword search on the Web only requires users to enter several keywords, which is very user-friendly. However, it cannot allow users to specify complex search conditions like graphs (such as relationships among keywords), and it only returns the Web hyperlinks that might contain answers to users. Hence, this simpleness also brings the gap between what the users want and what the users get. In contrast, the results of graph search  are much more accurate as it allows users to further specify  structural constraints by designing various pattern graphs. However, it is definitely inconvenient for users to enter pattern graphs as inputs even for small pattern graphs, as it is hard for non-professional users who are not familiar with the complex data graphs to specify precise pattern graphs.

People are already making an effort for designing friendly graph search interfaces. The technique developed by Facebook allows users to specify pattern graphs with simple natural language sentences, as we mentioned earlier. And Yang et al.~\cite{YangWuSY14} have recently proposed a novel graph search system enabling schemaless and structureless graph querying, which (a) provides a user-friendly interface where users can give rough descriptive pattern graphs as queries, and  (b) supports various kinds of transformations such as synonym, abbreviation, and ontology. However, a completely friendly interface that can meet the requirements of practical applications is still on its road for big graph search.

\stitle{(2) Accuracy}. It is necessary for a search engine to provide the users with accurate answers.

When a user submits a query to a search engine, which represents the user's searching goal, the search engine analyzes the user's input and tries to understand what the user wants. Hence, to reach high searching accuracy, it is indispensable to understand the users' real intents for search engines. However, it is pretty common that there is a gap between what a user wants and what she/he gets back from a search engine.
This is because it is a very challenging task to understand and specify the users' intents in a way such that a machine could easily understand. For example, when a user submits ``{\em apple}'' to a search engine, it is hard to distinguish the fruit apple from the products of Apple Inc..

Common approaches~\cite{StevenEric05,ShenSun06} focus on query classification. Given a query, these approaches try to classify the query to some predefined classes. Recently, some researchers take into account of the difference of individuals and attempt to analyze the intents of users by incorporating their search behaviors and preferences~\cite{XingLiu13,HuZhang11}.

Knowledge also plays an important role to understand the user intent and to improve the searching accuracy. For example, knowledge graph makes Google search engine more intelligent to understand the searching intents of users. When having the keyword ``apple'' into Google search engine, it will provide two extra panels in addition to a list of Web hyperlinks, one for Apple Inc. and the other for the apple fruit. Then users can click one to enlarge and get detailed information based on their intents, which allows users to get more relevant results without having to visit other Web sites to judge whether the information are relevant by themselves. This is because Google now is able to understand the difference among these entities, and the nuance in their meanings, with the aid of Knowledge Graph~\cite{GoogleKGraph2}.

\stitle{(3) Efficiency}. How to search information in a fast way is a key for the success of a search engine. It is also a fundamental problem in database and information retrieval areas, especially when we are dealing with big graphs today. We will introduce several searching techniques for big graphs in detail in the coming Section~\ref{sec-tech}.

\subsection{The Challenges}
\label{subsec-chellenges}

The expressiveness of graphs naturally comes with more difficulties, and the emerging social applications raise more challenges to search and manage big graphs.

According to statistics, for Facebook, there are over 1.3 billion monthly active users; for every 20 minutes, there are 1 million links shared, 2 million friend requests generated, and 3 million messages sent~\cite{facebooknum}; similarly for Twitter, there are over 0.6 billion users; every second there are 9100 tweets happened; and people query twitter search engine 2.1 billion times every day~\cite{twitterstatic}.

These statistics show the following. (a) Graph data have reached hundred millions orders of magnitude~\cite{massivegraph}; (b) Graph data are updated all the time, and the update amount daily reaches hundred thousands orders of magnitude~\cite{Albert06}; And, even worse, (c) Similar to traditional relational data~\cite{ErhardRahm00,vldbjFanLMTY12}, graph data have the data uncertainty problem due to the external reason caused by data sampling and data missing and the internal reason caused by the dynamic changes in graph data. In summary, graph data have three significant features: {\em big, dynamic and uncertain}~\cite{cccf-2}.
The first feature requires that graph search needs to strike a balance between its time and space cost. When the graph is too large to be processed on single machines, it is also necessary to design efficient and effective distributed algorithms.
The second feature requires that graph search should take dynamic changes and temporal factors into consideration.
%
%
The last feature requires that graph search should design reasonable models to capture uncertainties in graph data, and design highly efficient algorithms to answer graph search queries on uncertain graphs.

These together make it an extremely challenging task to develop a big graph search engine with a friendly query interface, accurate answers and high efficiency.

\eat{For example, Google's Knowledge Graph is a semantic network containing over 570 million objects, and more than 18 billion facts about the relationships between these different objects~\cite{GoogleKGraph1}. It is introduced into Google search engine to understand the meaning of the keywords entered for the search and promote the quality of searching results. With Knowledge Graph, Google search engine is enhanced from three aspects~\cite{GoogleKGraph2}: it (a) resolves the query ambiguity problem to help users find the right topic; (b) gets the best summary around the query topic to help understand the topic comprehensively and thoroughly; and (c) provides related or similar topics to make the search go deeper and broader. It is worth mentioning that in the first aspect, Knowledge Graph helps Google search engine better understanding the searching intents of users, which is an important issue in graph search we will discuss later. All these prove that graph search is a promising searching paradigm for social computing, and will play an important role in big data era.
}

\eat{
\stitle{Problems:} Currently, the study of keyword search on graphs mainly pays attention to query efficiency and the topology of query results, concerning less on the rationality of the semantic. The main reason is that it is extremely difficult to define a precise and universal query semantic for keyword search. In comparison, the query results are much more accurate via graph pattern matching. However, it is inconvenient to deal with the requirement of entering a pattern graph completely. Then a common idea arises: is it possible to propose a novel method based on "keyword search on graphs" and "graph pattern matching", which combines both advantages and overcomes shortcomings respectively? It remains to be an open question.\par
}

\vspace{-0ex}
\section{Techniques Towards Big Graph Search}
\label{sec-tech}

A fundamental issue in the big data era is the efficiency. In this section, we present three classes of techniques for big graph search: query techniques, data techniques and distributed computing techniques.

\subsection{Query Techniques}

 We first introduce two query techniques: query approximation and incremental computation.

\begin{figure}[h]
  \vspace{-1ex}
  \begin{center}
  \includegraphics[scale=0.45]{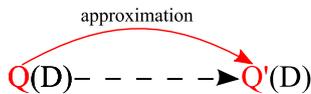}\\
  \end{center}
  \vspace{-3ex}
  \caption{Query approximation}\label{fig-tech-queryappro}
  \vspace{-2ex}
\end{figure}

\stitle{(1) Query approximation.} The core idea of query approximation is to transform a class of queries $Q$ with higher computational complexity into another class of queries $Q'$ with lower computational complexity and satisfiable approximate answers, as depicted in Fig.~\ref{fig-tech-queryappro} in which $Q$, $Q'$ and $D$ denote the original query, approximate query and data, respectively. The major challenge comes from the need of a balance between the query efficiency and answer accuracy.

We next explain the query approximation technique using {\em strong simulation}, a new graph pattern matching model proposed in~\cite{tods-MaCFHW14,ShuaiMaVLDB12}.
Graph pattern matching is to find all matched subgraphs in a data graph for a given pattern graph, and it is often defined in terms of {\em subgraph isomorphism}. The goodness of subgraph isomorphism is that all matched subgraphs  are exactly the same as the pattern graph, \ie completely preserving the  topology structure between the pattern graph and data graph.

Subgraph isomorphism is, however, \NP-complete~\cite{Ullmann76}, and may return exponential many matched subgraphs.
Recent evidences have shown that subgraph isomorphism is too restrictive to find sensible matches in certain scenarios~\cite{FanLMTWW10}. These hinder the usability of graph pattern matching in emerging applications.

To lower the high complexity of subgraph isomorphism, various extensions of graph simulation~\cite{infsimu95} have been considered instead in~\cite{FanLMTWW10,FanLMTW11}. These extensions allow graph pattern matching to be conducted in cubic-time. However, they fall short of capturing the topology of data graphs, i.e., graphs may have a structure drastically different from pattern graphs they match, and the matches found are often too large to analyze.

To rectify these problems, strong simulation, a revision of graph simulation, was  proposed for graph pattern matching, such that strong simulation (a) preserves the topology of pattern graphs and finds a bounded number of matches, (b) retains the same complexity as earlier extensions of graph simulation~\cite{FanLMTWW10,FanLMTW11}, by providing a cubic-time algorithm for computing strong simulation, and (c) has the locality property that allows us to develop an effective distributed algorithm to conduct graph pattern matching on distributed graphs~\cite{tods-MaCFHW14,ShuaiMaVLDB12}.

\eat{
Take the expertise recommendation network in Fig.~\ref{fig-recommendation} for example. When input pattern graph and data graph are $Q$ and $G$, and the ``match'' semantic is subgraph isomorphism, the search result shows that $Q$ and $G$ are not matched via subgraph isomorphism, \ie there exists no subgraphs in $G$ have exactly the same structural topology with $Q$.

Graph simulation looses the structural constraint of subgraph isomorphism, and can be solved in quadratic time. It does not require the match relation to be node-to-node bijection, but a binary relation between pattern nodes and matched nodes is enough that can preserve child relationship. In the recommendation example, the entire data graph $G$ is included in the match relation, which maps $HR$, $SE$,$Bio$, $DM$ and $AI$ in $Q$ to \{$HR_1$, $HR_2$\}, \{$SE_1$, $SE_2$\}, \{$Bio_1$, $Bio_2$, $Bio_3$, $Bio_4$\}, \{$DM'_1$, $DM'_2$, $DM_1$, ..., $DM_k$\} and \{$AI'_1$, $AI'_2$, $AI_1$, ..., $AI_k$\} in $G$, respectively. However, although graph simulation can be conducted in quadratic time, it falls short of capturing the topology of data graphs. As in the example, the matched graph has a structure drastically different from pattern graph.

Strong simulation enhances the expressive power of match semantic graph simulation by requiring the binary match relation to preserve both parent and child relationships, and also proposing the locality property ensuring the cohesive relations among matched nodes. In the recommendation example, the connected component composed of $HR_2$, $SE_2$, $Bio_4$, $DM'_1$, $DM'_2$, $AI'_1$ and $AI'_2$ is the match. Obviously, the cubic-time complexity of strong simulation is higher than the quadratic-time complexity of graph simulation, but strong simulation is better than graph simulation at preserving the topology of pattern graphs and data graphs, which strikes a balance between expressiveness and complexity. We can also find that although the structural preservation of strong simulation is not as strong as subgraph isomorphism, through the query semantic approximation, strong simulation can get reasonable match results in cubit time \wrt the \NP-completeness of subgraph isomorphism.
}

\eat{

such that the searching approach is inefficient, and it requires there exist subgraphs in the matched data graph have exactly the same structural topology with pattern graph, Recently, some approaches are proposed to convert the high complexity query semantic subgraph isomorphism to new query semantics with lower complexity, such as graph simulation and strong simulation.

 Apparently, the match semantic
keeps exact structure topology between Q and Gs.

To lower its complexity, various extensions of graph simulation have been considered instead~ . These extensions allow graph pattern matching to be conducted in cubic-time. However, they fall short of capturing the topology of data graphs, i.e., graphs
may have a structure drastically different from pattern graphs they match, and the matches found are of-
ten too large to understand and analyze.

Subgraph isomorphism is

introduction on the technique by an application: Graph pattern matching.

 According to the definition of Graph search in Section \ref{sec-definition}, graph pattern matching problem~\cite{FanLMTWW10,FanLMTW11} is given a pattern graph and a data graph, to identify all subgraphs ``match'' with pattern graph in data graph. Different ``match'' semantics form different graph pattern matching languages, such as subgraph isomorphism, graph simulation~\cite{infsimu95} and strong simulation~\cite{tods-MaCFHW14,ShuaiMaVLDB12}. We will next make an analysis on them and find how strong simulation query approximates subgraph isomorphism query.

\noindent{\em Graph isomorphism}. To introduce subgraph isomorphism, we first introduce graph isomorphism. Given a pattern graph $Q$ and a data graph $G$, $Q$ and $G$ are isomorphic if and only if there exists a bijection between the node sets of $Q$ and $G$ $f$:$V_Q\rightarrow V_G$, such that any two nodes $u$ and $v$ are adjacent in $Q$ if and only if $f(u)$ and $f(v)$ are adjacent in $G$.

\noindent{\em Subgraph isomorphism}. Given a pattern graph $Q$ and a data graph $G$, $Q$ and $G$ are matched via subgraph isomorphism if and only if there exists a subgraph $G_s$ in $G$ isomorphic with $Q$. Apparently, the match semantic keeps exact structure topology between $Q$ and $G_s$.
}

\begin{figure}[h]
\vspace{-1ex}
  \begin{center}
  \includegraphics[scale=0.45]{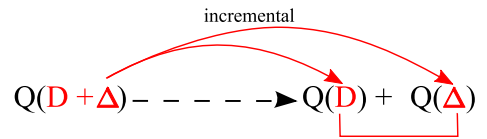}\\
  \end{center}
  \vspace{-3ex}
  \caption{Incremental computation}\label{fig-tech-queryincrem}
  \vspace{-2ex}
\end{figure}

\stitle{(2) Incremental computation}. When there are data updates, query answers typically need to be re-computed to reflect the changes. In practice, big data graphs are frequently modified, as we pointed out in Section~\ref{subsec-chellenges}, and it is too costly to recompute matches from scratch every time when the data graphs are updated. Incremental computation is a technique that attempts to reduce time by reusing previous computing efforts and only computing those answers that ``depend on'' the changed data, and it is depicted in Fig.~\ref{fig-tech-queryincrem}, in which $Q$, $D$ and $\Delta$ denote the query,  original data and its updates, respectively.

\eat{When a user inputs a query $Q$ into a search engine with data set $D$ already stored in, and gets the query result $Q(D)$. If the data set is time evolved and after a period of time, there exist changes $\Delta$ on $D$. If the user wants to find the new query result of $Q$ on $D+\Delta$, however, it is expensive to recompute the result from scratch via batch algorithm due to the slight changes in data set, and this comes the convenience for executing incremental computing. It just computes changes $Q(\Delta)$ based on previous query result corresponding to the updates in data set, and then gets the new query result by merging $Q(\Delta)$ with the previous one $Q(D)$, as depicted in Fig.~\ref{fig-tech-queryincrem}.
}

It is worth mentioning that incremental algorithms have been developed for
various applications (see~\cite{inc-survey} for a survey).  Thomas W. Reps has done pioneering work on the study of incremental computation~\cite{inc-survey,Reps96}, and he observed in~\cite{Reps96} that
the complexity of an incremental algorithm is more accurately characterized
in terms of the size of the area affected by the updates, rather than the size
of the entire input.

Next let's take the indexing of Google search as an example. It is known that the Web documents are crawled and stored in a large repository, and are pre-indexed to speed up the search efficiency and improve the user experiences. The indexing process incurs a heavy workload, and Google initially adopted some batch-processing approaches such as MapReduce~\cite{DeanG04} to improve the efficiency, which is not satisfactory when facing with constant changes.
Google later on developed Percolator~\cite{Daniel10}, a system incrementally processing updates on large data sets. That is, Google has converted its  batch-based indexing system into an incremental indexing system. It was reported that compared with MapReduce, Percolator (a) reduced the average document processing latency by a factor of 100, and (b) reduced the average age of resulting documents of Google search by 50\% when processing the same amount of documents per day~\cite{Daniel10}.

\subsection{Data Techniques}

One key feature of big data graphs is the large volume, and, hence, the space complexity~\cite{Papa1994} of graph search starts raising more troubles. Here we introduce five techniques to  boost the search efficiency from the data point of view: data approximation, data sampling, data partitioning, data compression and data indexing.

\begin{figure}[h]
  \vspace{-1ex}
  \begin{center}
  \includegraphics[scale=0.45]{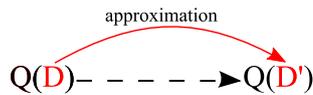}\\
  \end{center}
  \vspace{-3ex}
  \caption{Data approximation}\label{fig-tech-dataappro}
  \vspace{-2ex}
\end{figure}

\stitle{(1) Data approximation}. The core idea of data approximation is that given a class of queries $Q$ and a data set $D$, it transforms $D$ into a smaller data set $D'$ such that $Q$ on $D'$ returns a satisfiable approximate answer in a more efficient way, as depicted in Fig.~\ref{fig-tech-dataappro}.
Similar to query approximation, the major challenge of data approximation comes from the need of a balance between the query efficiency and answer accuracy.

We have adopted the idea in the process of dealing with large graphs in the study of anomaly detection in graph streams, when dealing with the matrix representation of a social graph, and  we have both theoretically and experimentally shown that simplifying the matrix by replacing a part of small entry values  with zero has few affects on the computation of eigenvectors~\cite{YuAMW13}.

\eat{In another study of anomaly detection in community detection, suppose a social graph is composed of $d$ communities and we want to represent the person-community relationship with matrix. Each node represents a person and can be embedded into $d$-dimensional space, in which each dimension indicates the membership of this node to the corresponding community. Thus the graph can be represented as a $n*d$ matrix. Sometimes the value $d$ may be rather large, therefore the space and computation complexity of maintaining and operating on the matrix are relatively high. Note that each node only has direct interactions with a limited number of communities, rather than the whole graph, which inspires us to simplify each $d$-dimensional vector to top $k$ dominant dimensions, and the rest dimensions are simply eliminated. Here $k$ is the maximum number of communities that each node can belong to. These two approximate techniques can reduce both space and time costs while introduces little loss of quality, as shown by the experimental study.}

\begin{figure}[h]
  \vspace{-1ex}
  \begin{center}
  \includegraphics[scale=0.45]{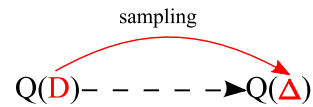}\\
  \end{center}
  \vspace{-3ex}
  \caption{Data sampling}\label{fig-tech-datasampl}
  \vspace{-2ex}
\end{figure}

\stitle{(2) Data sampling}. Sampling is concerned with the selection of a subset of data from a large data set. Instead of dealing with the entire data set $D$ for a query $Q$, the data sampling technique reduces the size of the data set $D$ by sampling, with a permission of loss of accuracy to some extent in the query result~\cite{Aggarwal10}. In a sampling process, it must be ensured that the sampled data $\Delta$ obtained must reflect the characteristics and information of the original data $D$, as depicted in Fig.~\ref{fig-tech-datasampl}.

It is worth mentioning that Michael I. Jordan and his colleagues have proposed a new sampling approach --bootstrap-- to dealing with big data~\cite{Jordan12,KleinerTSJ12}.

\begin{figure}[h]
  \vspace{-1ex}
  \begin{center}
  \includegraphics[scale=0.45]{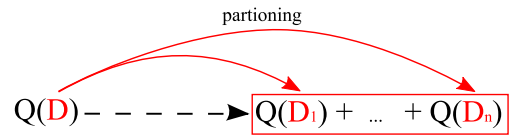}\\
  \end{center}
  \vspace{-3ex}
  \caption{Data partitioning}\label{fig-tech-datapart}
  \vspace{-2ex}
\end{figure}

\stitle{(3) Data partitioning}.  Data partitioning is an effective method to execute queries on large-scale data sets in a divide-and-conquer way. It partitions a data set $D$ into a set of {\em relatively small} data sets $D_1$, $\cdots$, $D_n$ such that $D = D_1\cup\cdots\cup D_n$. Ideally, the final query answer is assembled using the $n$ answers on the set of small data sets, and the analysis speed can be improved significantly. The entire process is depicted in Fig.~\ref{fig-tech-datapart}.

It is worth mentioning that graph partitioning has been extensively studied since the 1970's~\cite{kl70,Karypis98,YangYZK12},
and has been successfully used in various applications, \eg circuit placement, parallel computing and scientific simulation~\cite{YangYZK12}.
The graph partitioning problem is in general a hard problem and is often \NP-complete~\cite{Karypis98}.

\stitle{(4) Data compression}. The principle of data compression is that compressing by removing redundancies
also answers the same question. There are many known data compression methods that are suitable
for different types of data, and produce different answers, but they are all based on the principle, namely
compressing data by removing redundancies from the original data (see~\cite{compressionbook} for a complete reference).
The benefits of data compression lie in that it provides more possibilities to work in main memory and potentials to work efficiently.

\begin{figure}[h]
 \vspace{-1ex}
  \begin{center}
  \includegraphics[scale=0.45]{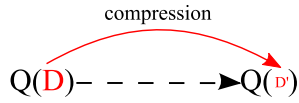}\\
  \end{center}
  \vspace{-3ex}
  \caption{Data compression}\label{fig-tech-datacompression}
  \vspace{-2ex}
\end{figure}

Different from data sampling, data compression generates a small data set $D'$ from the original data set $D$ by removing redundancies and preserving the information only relevant to queries, as depicted in Fig.~\ref{fig-tech-datacompression}. In addition, there are usually no restrictions on the formats of the compressed data, while data sampling normally keeps the original data formats. There is a whole bunch of work on (lossy or lossless) graph compression~\cite{BuehrerC08,AdlerM01,BoldiV04,MaserratP10}. As~\cite{FederM95, KarandeCA09,FanLiWang12} show, some graph algorithms can be speeded up by operating on compressed graphs directly, which can be treated as query oriented compression, and needs to invest more efforts to study.

\stitle{(5) Data indexing}. An index is a data structure that improves the speed of queries by reducing search space, at the cost of update maintenance and extra storage. Indexes are commonly used for querying relational databases~\cite{Raghu2000} and information retrieval of search engines~\cite{irbook}.

When data graphs are relatively large, graph indexing technique can quickly prune data graphs that obviously mismatch the pattern graph~\cite{KleinKM11}. There already exist indexing methods for (various kinds of) graph pattern matching~\cite{Aggarwal10}. There are mainly three metrics for measuring whether an established index  is appropriate: the space cost, building time and query time. The smaller the space of an index is, the less additional storage burden incurred. The building time represents the time cost of creating the index, and the query time indicates the time cost for the query process. When data graphs are changed over time, the index refreshing speed represents its ability to adapt to dynamic changes.

\subsection{Distributed Computing Techniques}
We now introduce the distributed computing techinque that utilizes the query and data techniques and beyond.

\begin{figure}[h]
 \vspace{-1ex}
  \begin{center}
  \includegraphics[scale=0.45]{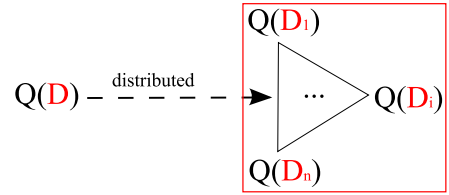}\\
  \end{center}
  \vspace{-3ex}
  \caption{Distributed computing}\label{fig-tech-disproc}
  \vspace{-2ex}
\end{figure}

Distributed computing refers to the use of distributed systems to solve problems such that a problem is divided into many tasks, each of which is computed on one or more machines, and which communicate with each other by message passing~\cite{dabook,dcbook}. Distributed computing typically needs to partition a data set $D$ into {\em relatively small} data sets $D_1$, $\cdots$, $D_n$, and distributes them on multiple computing machines, as depicted in Fig.~\ref{fig-tech-disproc}.

It is known that real-life graphs are typically way too large, \eg the Web graph of Yahoo! has about 14 billion nodes, and there are over 1.3 billion users on Facebook. Hence, it is not practical to handle large graphs on single machines. Moreover, real-life graphs are naturally distributed, \eg Google, Yahoo! and Facebook have large-scale distributed data centers. This says that distributed computing is inevitable facing with big graphs.

We have developed a computation model for a large class of distributed algorithms for graph simulation \cite{ShuaiMaWWW12}. The model consists of a cluster of identical machines, in which one acts as a coordinator. Each machine can directly send an arbitrary number of messages to another, and all machines co-work with each other by local computations and message-passing. Further, we also identify three complexity measures on the performance of distributed algorithms related to the computation model above: (a) visit times, which is the maximum visiting times of a machine, indicates the complexity of interactions; (b) makespan, which is the evaluation of the total computation time, is a measure of efficiency; (c) data shipment, which is the size of the total messages shipped among distinct machines during the computation, indicates the network bandwidth consumption. However, these three measures are typically controversial with each other, and how to achieve a balanced strategy is a great challenge for designing distributed algorithms.

Recently, many distributed graph processing systems have been developed, which basically fall into two categories: one makes use of MapReduce~\cite{DeanG04}  or Spark~\cite{Spark} to speed-up big graph processing~\cite{GaoZZY14,QinYuxu14,GraphX}, and the other uses different distributed computing models, such as  Pregel~\cite{MalewiczABDHLC10}, GraphLab~\cite{GraphLab12} and PowerGraph~\cite{PowerGraph12}.

\eat{
MapReduce~\cite{DeanG04} is a programming model that allows for massive scalability across a large cluster of computing machines. Pregel~\cite{MalewAust10} is a distributed programming framework suitable for large-scale graph computing with high efficiency and scalability. Researchers have defined \texttt{MMC} for algorithm classes in MapReduce in terms of memory consumption, communication cost, CPU cost, and son on. Qin et al.~\cite{QinYuxu14} have defined a new class \texttt{SGC} for scalable graph processing in MapReduce, extending the definition of \texttt{MMC}. They have also defined two graph operators which can be used to design a large range of graph algorithms.
}

\eat{
\, It is known that query efficiency can be improved by conducting a preprocessing step on pattern graphs. There are mainly two kinds of methods, query minimization and query similarity transformation. The previous one generates a new pattern graph $Q'$ by removing redundant nodes and edges in the primary pattern graph $Q$, while ensuring they must return same query results on any data graphs. The second method will simplifies pattern graph $Q$ with a permission of loss of precision to some extent in the query result, such as to represent $Q$ as a combination of a set of pattern graphs \{$Q'_1, Q'_2,..., Q'_k$\} with known query results. Subsequently, by making use of the known query results of \{$Q'_1, Q'_2,..., Q'_k$\}, final query result of $Q$ can be derived. Concrete measures for query similarity transformation are generally produced based on the characteristics of specific graph query languages.\par
}

\stitle{Remarks}. There exist no single techniques that could fit all for big graph search. That is, it is often necessary to combine different techniques to obtain satisfiable solutions.   We also encourage interested readers to read a very recent article~\cite{jcstFanH14} for discussions on the theory and techniques of big data, a complement of this article.

\vspace{-0ex}
\section{Conclusions}
\label{sec-conclusion}

In this article we have investigated big graph search, a novel promising search paradigm for social computing in the big data era. First, we have analyzed the need of big graph search with various applications, industrial and academic developments, and the evolution history of information searching paradigms. Second, we have pointed out the challenges and opportunities of big graph search. Finally, we have introduced three types of techniques towards big graph search: query techniques, data techniques and distributed computing techniques.

Being a new paradigm for social computing, big graph search has received extensive attentions. However, there is obviously a long way to go for a  big graph search engine that meets various needs in practice.

\stitle{Acknowledgments}.
This work is supported in part by  973 program  ({\small No. 2014CB340300}), NSFC ({\small No. 61322207}) and the Fundamental Research Funds for the Central Universities.
\balance
\bibliographystyle{abbrv}
\begin{small}
\bibliography{paper}
\end{small}

\end{document}